\documentclass[letterpaper, 10 pt, conference]{ieeeconf}  

\IEEEoverridecommandlockouts                              
\usepackage[utf8]{inputenc}
\usepackage[T1]{fontenc}

\usepackage{graphicx,float,dblfloatfix} 
\usepackage{amsmath} 
\usepackage{amssymb}  
\usepackage{bm}
\usepackage[hidelinks]{hyperref}
\usepackage{algorithm}

\title{\LARGE \bf
Intelligent Control for Path-Following of an Unmanned Mass-Centric Surface Vehicle
}

\author{Gabriel S. Lima$^{1}$, Tomi Westerlund$^{2}$, and Wallace M. Bessa$^{1}$
\thanks{*This work was not supported by any organization.}
\thanks{$^{1}$G. S. Lima and W. M. Bessa are with the Smart Systems Lab, University of Turku, 20520 Turku, Finland (e-mail: gdasil@utu.fi, wmobes@utu.fi).}%
\thanks{$^{2}$T. Westerlund is with the Turku Intelligent Embedded and Robotic Systems Research Group, University of Turku, 20520 Turku, Finland (e-mail: tovewe@utu.fi).}%
}

\begin{document}

\maketitle
\thispagestyle{empty}
\pagestyle{empty}

\begin{abstract}

Addressing the control and maneuverability of surface vehicles with dynamically changing mass distributions is still an open problem. To solve the problem, we propose an intelligent controller for the path-following problem of a surface vehicle, which is controlled through mass distribution. This means that one of the control inputs is mass-centric. Specifically, we developed a Lyapunov-based nonlinear control scheme to enable an unmanned vessel to follow a smooth path according to a line-of-sight guidance law. The control inputs consist of the thrust force for forward motion and the position of a sliding mass that shifts the system’s overall mass distribution. Artificial neural networks are employed to estimate unmodeled dynamics and external disturbances. Simulation results demonstrate the effectiveness of the proposed controller in guiding the vessel along the desired path with minimal error.

\end{abstract}

\section{INTRODUCTION}

\begin{figure}[ht]
    \centering
    \includegraphics[width=0.7\linewidth]{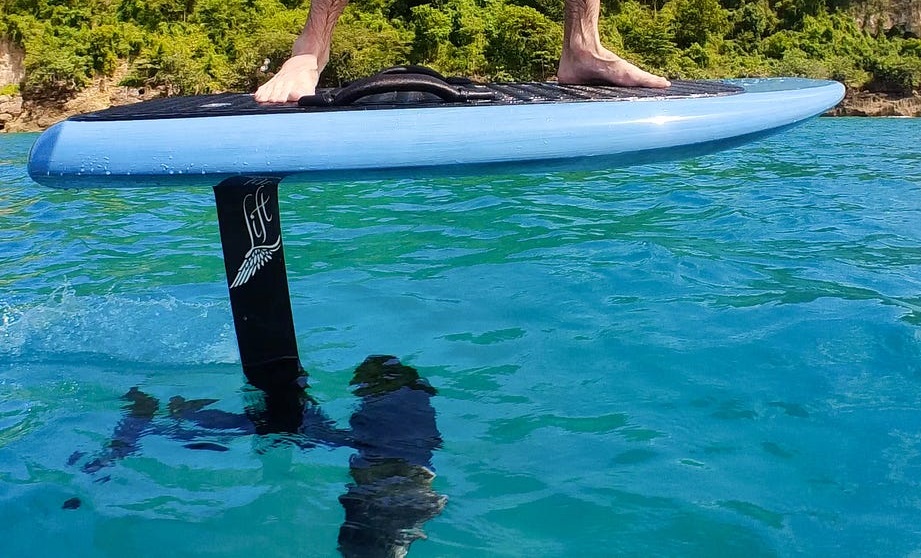}
    \caption{Electric hydrofoil board.}
    \label{fig:efoil}
\end{figure}

Efoils, or electric hydrofoil boards (Fig. \ref{fig:efoil}), are surfboard-like devices equipped with a hydrofoil and an electric motor that allows riders to "fly" above the water. They consist of a board, a hydrofoil wing extending below the water surface, and an electric propeller powered by a battery. This configuration enables efoils to lift out of the water as the board reaches certain speeds, giving the impression of gliding on air. Riders use a handheld remote to control the speed, making efoils relatively easy to operate, and are often seen using them for recreation in lakes, rivers, and coastal areas. Their quiet, motorized design attracts water sports enthusiasts looking for a unique, eco-friendly way to explore water bodies without creating noise or water pollution, which further sets efoils apart from traditional motorized watercraft.

Beyond recreational use, autonomous efoils show promising potential for applications in surveillance and monitoring of water bodies, particularly in big lakes and seas. Their low profile and minimal noise make them ideal for discreet observation, allowing researchers or authorities to monitor wildlife, water quality, or human activity without disturbing the environment. Autonomous efoils can be programmed to follow predetermined routes and gather real-time environmental data about the health of aquatic ecosystems. In addition, these self-navigating efoils can be equipped with cameras and sensors to detect anomalies like pollution or illegal fishing, making them valuable tools for environmental protection and waterway security.

Designing an autonomous efoil presents a unique challenge in determining how to control its direction. Typically, a rider steers the efoil simply by shifting their weight, which changes the board’s center of gravity to achieve the desired maneuver. In the absence of a rider, however, alternative methods must be employed to achieve this directional control. One approach is to equip the efoil with additional thrusters on each side, allowing for precise directional adjustments as has been demonstrated in~\cite{rctestflight_2024}. Another possibility involves using a sliding mass mechanism that shifts the center of mass of the system~\cite{bernad2019hydrofoil}. While this second option poses a more complex engineering challenge, it offers benefits in terms of installation simplicity and energy efficiency, as it requires only a single mechanical system powered by the existing battery.

Intelligent controllers play an essential role in autonomous vehicles, as these systems typically operate in changing and dynamic environments with high levels of uncertainty. Therefore, nonlinear controllers combined with computational intelligence schemes are an effective approach for controlling these systems, enabling them to perform tasks with minimal error~\cite{bessa2018biologically}. For instance, intelligent controllers have been applied for mobile robots~\cite{da2023accurate, ahmad2023modeling}, drones~\cite{xu2022identification, abdillah2024new}, robotic manipulators~\cite{lima2021intelligent, ren2022novel, sachan2024intelligent}, among other applications~\cite{lima2023intelligent,ren2023disturbance, el2024robust}.

With the aim of developing a control system for an efoil, this paper focuses on a simplified version of the system. We present the design of an intelligent controller for the path-following problem, considering only the planar motion of the system. Specifically, the system will have three degrees of freedom: two for translation and one for rotation. The primary framework of the controller is based on a Lyapunov-derived control law designed to ensure robustness within the system. To address uncertainties and nonlinearities, we employ a single hidden-layer artificial neural network. The control inputs consist of the thrust force generated by the propeller and the position of a sliding mass that adjusts the system's mass distribution. To determine the latter variable, we utilize an inverse dynamics approach, which calculates the position of the sliding mass from the torque necessary to move the system. The Line-of-Sight (LOS) algorithm~\cite{fossen2014line} will be used to estimate the desired heading angle.

\section{MATHEMATICAL MODEL}

\begin{figure}[ht]
    \centering
    \includegraphics[width=0.7\linewidth]{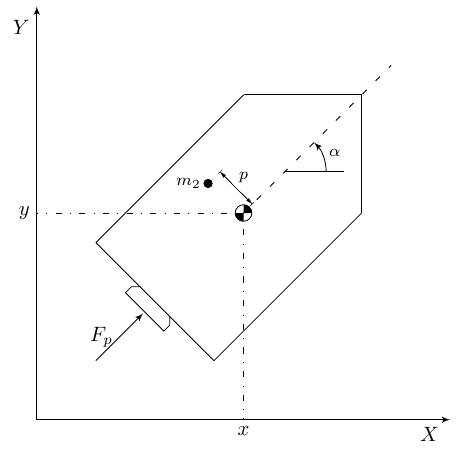}
    \caption{Top-view of the mass-centric vessel.}
    \label{fig:top}
\end{figure}

A top view of the vessel is illustrated in the Fig.~\ref{fig:top}. The system's position is represented by the vector $\bm{q} = [x,y,\alpha]^\top$. 

Using the Newton's Second Law, the equation of motion of the system can be written as follows:
\begin{equation}\label{eq:model0}
    \bm{M} \ddot{\bm{q}} = \bm{F} + \bm{u}
\end{equation}
in which the terms are described below:
\begin{equation}\bm{M} =
    \left[\begin{array}{ccc}
        (m_1 + m_2) c(\alpha)  &  (m_1 + m_2) s(\alpha) & -m_2 p\\
        -(m_1 + m_2) s(\alpha)  &  (m_1 + m_2) c(\alpha) & 0\\
         -m_2 p c(\alpha)  &  -m_2 p s(\alpha) & I_z + m_2 p^2
    \end{array} \right]
\end{equation}
\begin{equation}\bm{F} =
    \left[\begin{array}{c}
        f_{dx}\\
        m_2 p\dot{\alpha} + f_{dy}\\
         M_{dz}
    \end{array} \right]
\end{equation}
\begin{equation}\bm{u} =
    \left[\begin{array}{c}
        F_{p}\\
        0\\
         M_{a}
    \end{array} \right]
\end{equation}
where $m_1$ and $m_2$ are the masses of the vessel and the sliding mass, respectively; $I_z$ is the vessel's moment of inertia; $p$ is the position of the sliding mass relative to the vessel’s center; $f_{dx}$, $f_{dy}$, and $M_{dz}$ represent the hydrodynamic friction forces in each degree of freedom; $F_p$ is the thrust force; and $M_a$ is the torque required to rotate the vessel. Additionally, we define $s(\alpha) = \sin\alpha$ and $c(\alpha) = \cos\alpha$.

Converting the system dynamics to the body-fixed reference frame using the transformation $\bm{v} = \bm{T}_r \dot{\bm{q}}$, where $\bm{T}_r$ is the transformation matrix:
\begin{equation}\bm{T}_r =
    \left[\begin{array}{ccc}
        c(\alpha)  &  s(\alpha) & 0\\
        -s(\alpha)  &  c(\alpha) & 0\\
         0  &  0 & 1
    \end{array} \right]
\end{equation}
with $\bm{v} = [v_x,v_y,\dot{\alpha}]^\top$ being the speed vector, the system dynamics can be rewritten as follows:
\begin{equation}\label{eq:model1}
    \dot{\bm{v}} = \bar{\bm{F}} + \bm{B}\bm{u}
\end{equation}
where $\bar{\bm{F}} = \dot{\bm{T}}_r \dot{\bm{q}} + \bm{T}_r \bm{M}^{-1} \bm{F}$ and $\bm{B} = \bm{T}_r \bm{M}^{-1}$. The term $v_x$ is called surge speed.

Given that we will use the LOS algorithm to determine the desired heading angle for guiding the vessel, we focus only on the following equations to derive the control scheme:
\begin{align}
    \dot{v}_x = f_x + b_{x} F_p + d_x \label{eq:vx}\\
    \ddot{\alpha} = f_\alpha + b_{\alpha} M_a + d_\alpha \label{eq:a}
\end{align}
where $f_x = \bar{\bm{F}}_1$, $f_\alpha = \bar{\bm{F}}_3$, $b_x = \bm{B}_{11}$, and $b_\alpha = \bm{B}_{33}$. Additionally, the terms $d_x$ and $d_\alpha$ represent all modeling inaccuracies, external disturbances, and neglected dynamics, including the coupling terms from $\bm{B}_{13}$ and $\bm{B}_{31}$.

\section{CONTROLLER DESIGN}

Consider a path $\mathcal{P}(\theta)$ described in the plane and in function of a time-varying parameter $\theta$. Its coordinates are given by $x_d = x_\mathcal{P}(\theta)$ and $y_d = y_\mathcal{P}(\theta)$. The tracking error in the reference frame attached to the path $\mathcal{P}$ is given by:
\begin{equation}
    \left[\begin{array}{c}
        x_e\\
        y_e
    \end{array} \right] = \left[\begin{array}{cc}
        c({\alpha_\mathcal{P}})  &  -s({\alpha_\mathcal{P}}) \\
        s({\alpha_\mathcal{P}})  &  c({\alpha_\mathcal{P}})
    \end{array} \right] \left[\begin{array}{c}
        x - x_d\\
        y - y_d
    \end{array} \right]
\end{equation}
where $\alpha_\mathcal{P} = \arctan (y_d^\prime, x_d^\prime)$, with $y_d^\prime = \partial y_d /\partial \theta$ and $x_d^\prime = \partial x_d /\partial \theta$.

Then, following \cite{fossen2014line,yan2023continuous}, we can define the dynamics of $\theta$ and the desired heading angle by:
\begin{align}
    \dot{\theta} = \frac{k x_e + v_x\cos(\alpha - \alpha_\mathcal{P}) - v_y\sin(\alpha - \alpha_\mathcal{P})}{\sqrt{{x_d^\prime}^2 + {y_d^\prime}^2}} \label{eq:th}\\
    \alpha_d = \alpha_\mathcal{P} - \arctan\left( \frac{y_e}{\Delta}\right) - \beta
\end{align}
with $k > 0$ being a control parameter, $\Delta > 0$ the lookahead distance, and $\beta = \arctan(v_y,v_x)$ the sideslip angle.

Following the feedback linearization approach \cite{slotine1991applied}, we can define the controller for dynamics described by~\eqref{eq:vx} and~\eqref{eq:a}:
\begin{align}
    F_p = b_x^{-1} (-\hat{f}_x - \hat{d}_x + \dot{v}_{xd} - \lambda_x s_x) \label{eq:u-vx}\\
    M_a = b_\alpha^{-1} (-\hat{f}_\alpha - \hat{d}_\alpha + \ddot{\alpha}_{d} - \lambda_\alpha \dot{\tilde{\alpha}} - \lambda_\alpha s_\alpha) \label{eq:u-a}
\end{align}
where $\hat{f}_x$, $\hat{f}_\alpha$, $\hat{d}_x$, and $\hat{d}_\alpha$ are estimates of  $f_x$, $f_\alpha$, $d_x$, and $d_\alpha$, respectively, with the uncertainty related to $f_x$ and $f_\alpha$ being included into $d_x$ and $d_\alpha$. Also, we have the variables inspired by the sliding control method $s_x = v_x - v_{xd}$ and $s_\alpha = \dot{\tilde{\alpha}} + \lambda_\alpha \tilde{\alpha}$ with $\tilde{\alpha} = \alpha - \alpha_d$ being the tracking error. The terms $\lambda_x > 0$ and $\lambda_\alpha > 0$ are control parameters.

\begin{figure}[ht]
    \centering
    \includegraphics[width=0.65\columnwidth]{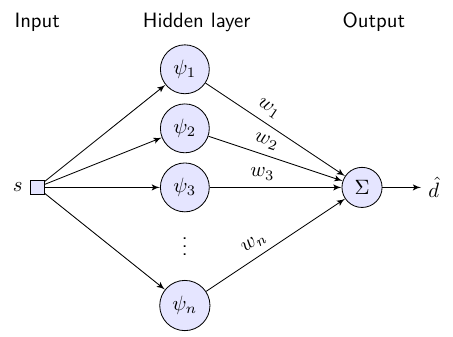}
    \caption{Single-hidden layer network.}\label{fig:ann}
\end{figure}

So, consider that a single-hidden layer artificial neural network, as depicted in Fig.~\ref{fig:ann}, can perform universal approximation~\cite{scarselli1998universal} and approximate the unknown dynamics $d_i$:
\begin{equation}
    \hat{d}_i = \bm{w}_i^\top \bm{\psi}_i(s_i)
\end{equation}
where $\bm{w}_i = [w_{i1} ~\ldots~ w_{in}]^\top$ are the weights of the neural network and $\bm{\psi}_i(s_i) = [\psi_{i1} ~\ldots~ \psi_{in}]^\top$ the activation functions with $i=x,\alpha$. 

Admitting there is a vector of optimum weights $\bar{\bm{w}}_x$ that minimizes the approximation error $\epsilon_x = d_x - \bar{\bm{w}}_x^\top \bm{\psi}_x$, where $\vert\epsilon_x\vert \leq \varepsilon_x$, we can define a candidate Lyapunov function as follows:
\begin{equation}
    V_x(t) = \frac{1}{2} s_x^2 + \frac{1}{2 \eta_x} \tilde{\bm{w}}_x^\top \tilde{\bm{w}}_x
\end{equation}
where $\eta_x$ is a strictly positive constant and $\tilde{\bm{w}}_x = \bar{\bm{w}}_x - \bm{w}_x$. 

Computing the time derivative of $V_x(t)$, considering that $\dot{\tilde{\bm{w}}}_x = - \dot{\bm{w}}_x$, and applying~\eqref{eq:u-vx}, we have
\begin{align}
    \dot{V}(t) & = s_x \dot{s}_x - \eta_x^{-1} \tilde{\bm{w}}_x^\top \dot{\bm{w}}_x = s_x [\dot{v}_x - \dot{v}_{xd}] - \eta_x^{-1} \tilde{\bm{w}}_x^\top \dot{\bm{w}}_x \nonumber \\
    & = s_x [f_x + b_{x} F_p + d_x - \dot{v}_{xd}] - \eta_x^{-1} \tilde{\bm{w}}_x^\top \dot{\bm{w}}_x \nonumber \\
    & = s_x [d_x - \hat{d}_x - \lambda_x s_x] - \eta_x^{-1} \tilde{\bm{w}}_x^\top \dot{\bm{w}}_x \nonumber \\
    & = s_x [\bar{\bm{w}}_x^\top \bm{\psi}_x - \bm{w}_x^\top \bm{\psi}_x + \epsilon_x - \lambda_x s_x] - \eta_x^{-1} \tilde{\bm{w}}_x^\top \dot{\bm{w}}_x \nonumber \\
    & = s_x [\tilde{\bm{w}}_x^\top \bm{\psi}_x + \epsilon_x - \lambda_x s_x] - \eta_x^{-1} \tilde{\bm{w}}_x^\top \dot{\bm{w}}_x \nonumber \\
    & = s_x [\epsilon_x - \lambda_x s_x] - \eta_x^{-1} \tilde{\bm{w}}_x^\top [ \dot{\bm{w}}_x - \eta_x s_x \bm{\psi}_x]
\end{align}

By definition of the adaptation law for the neural weights for the dynamics of $v_x$ as
\begin{equation}\label{eq:wpx}
    \dot{\bm{w}}_x = \eta_x s_x \bm{\psi}_x
\end{equation}
the first derivative of $V_x (t)$ becomes,
\begin{equation}\label{eq:Vpx}
    \dot{V}_x(t) = - [\lambda_x s_x - \epsilon_x ] s_x \leq - [\lambda_x \vert s_x \vert - \varepsilon_x ] \vert s_x \vert
\end{equation}

The exact same procedure can be followed to $s_\alpha$. Therefore, defining a candidate Lyapunov function as $V_\alpha(t) = \frac{1}{2} s_\alpha^2 + \frac{1}{2 \eta_\alpha} \tilde{\bm{w}}_\alpha^\top$, we can derive the adaptation law for the neural weights for the dynamics of $\alpha$ as
\begin{equation}\label{eq:wpa}
    \dot{\bm{w}}_\alpha = \eta_\alpha s_\alpha \bm{\psi}_\alpha
\end{equation}
So, the first derivative of $V_\alpha (t)$ becomes,
\begin{equation}\label{eq:Vpa}
    \dot{V}_\alpha(t) = - [\lambda_\alpha s_\alpha - \epsilon_\alpha ] s_\alpha \leq - [\lambda_\alpha \vert s_\alpha \vert - \varepsilon_\alpha ] \vert s_\alpha \vert
\end{equation}

It should be noted that equations~\eqref{eq:Vpx} and~\eqref{eq:Vpa} do not guarantee that \( \Vert \bm{w}_i \Vert_2 \) will be bounded when \( \vert s_i \vert \leq \varepsilon_i/\lambda_i \), with $i \in \{x,\alpha\}$. To address this issue, the projection algorithm~\cite{ioannou2006adaptive} can be used to ensure that \( \bm{w}_i \) will remain within the region \( \mathcal{W}_i = \{\bm{w}_i \in \mathbb{R}^n : \bm{w}_i^\top \bm{w}_i \leq \mu_i^2 \} \):
\begin{equation}\label{eq:projection}
    \dot{\bm{w}}_i = \left\lbrace \begin{array}{cl}
        \eta_i s_i \bm{\psi}_i &\quad\text{if}~\Vert \bm{w}_i \Vert_2 < \mu_i ~\text{or}  \\ 
         & \quad \text{if}~(\Vert \bm{w_i} \Vert_2 = \mu_i ~\text{and}~ \\
         & \quad \eta_i s_i \bm{w}_i^\top \bm{\psi}_i < 0) \\
         \left( \bm{I} - \frac{\bm{w}_i \bm{w}_i^\top}{\bm{w}_i^\top \bm{w}_i} \right)\eta_i s_i \bm{\psi}_i  &\quad\text{otherwise}
    \end{array} \right.
\end{equation}
where $\mu_i$ is the desired upper limit of $\Vert \bm{w}_i \Vert_2$ and $\eta_i$ will be called the learning rate of the neural network.

Since \( \Vert \bm{w}_i(0) \Vert_2 \leq \mu_i \), it follows that \( \vert s_i \vert \leq \varepsilon_i/\lambda_i \) and that \( \Vert \bm{w}_i \Vert_2 \leq \mu_i \) as \( t \rightarrow \infty \), which guarantees that the controller will ensure the exponential convergence of the tracking error to a closed region \cite{bessa2009some}, with $i=x,\alpha$.

It follows from \cite{yan2023continuous} that as $s_x, s_\alpha \rightarrow 0$, $\vert x_e \vert \rightarrow 0$, and $\vert y_e \vert$ is ultimately bounded by $\vert y_e \vert \leq \gamma \Delta$, where $\gamma$ is a positive constant related to the control parameters.

\begin{figure}[hb]
    \centering
    \includegraphics[width=\columnwidth]{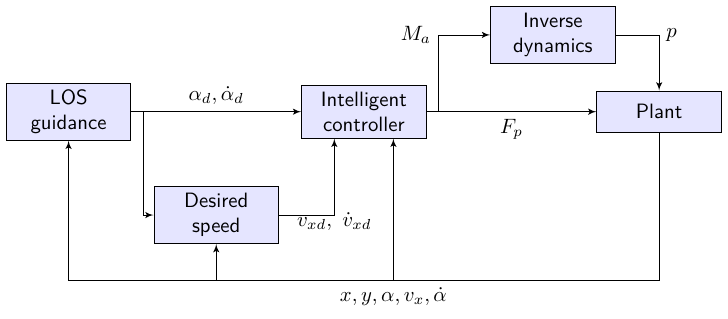}
    \caption{Controller framework.}\label{fig:frame}
\end{figure}

Finally, since the position $p$ of the sliding mass is our actual control parameter for changing the vessel's direction, we can calculate $p$ by isolating it in~\eqref{eq:model0}, using the computed value of $M_a$ from~\eqref{eq:u-a}. The control scheme is illustrated in the Fig.~\ref{fig:frame}.

\section{SIMULATION RESULTS}

The simulation was implemented in Python with the model \eqref{eq:model0} integrated using the Runge-Kutta method with 50\,Hz of sampling rate. The following model parameters were used in the simulations: $m_1 = 150$\,kg, $m_2 = 3$\,kg, and $I_z = 35.63$\,kg$\cdot$m$^2$. The hydrodynamic forces were defined as $f_{dx} = C_{dx} \dot{\bar{x}} \vert \dot{\bar{x}} \vert$, $f_{dy} = C_{dy} \dot{\bar{y}} \vert \dot{\bar{y}} \vert$, and $M_{dz} = C_{dz} \dot{\alpha} \vert \dot{\alpha} \vert$, where $\dot{\bar{x}} = \dot{x} - \dot{x}^\ast$ and $\dot{\bar{y}} = \dot{y} - \dot{y}^\ast$ are the speed relative to the water stream ($\dot{x}^\ast,\dot{y}^\ast$). The hydrodynamic coefficients used were $C_{dx} = 6.5$\,kg/m, $C_{dy} = 11$\,kg/m, and $C_{dz} = 0.4\,$kg$\cdot$m$^2$. Regarding the control parameters, we adopted $\lambda_x = 0.5$, $\lambda_\alpha = 0.4$, $k = 1.5$, and $\Delta = 200$\,m. It should be noted that the hydrodynamic forces are not considered in the controller, and a 20\% error was introduced in the masses. For the artificial neural network (ANN), Gaussian activation functions were used: $\psi_{ij} = \exp\{-0.5[(s_i - c_j)/l_j]^2\}$ with centers and widths being, respectively, $\bm{c}_i = [-\phi_i, -\phi_i/2, -\phi_i/4, 0, \phi_i/4, \phi_i/2, \phi_i]^\top$ and $\bm{l}_i = [\phi_i/2, \phi_i/3, \phi_i/4, \phi_i/4, \phi_i/4, \phi_i/3, \phi_i/2]^\top$, where $\phi_x = 2$ and $\phi_\alpha = 0.1$. The learning rates used were $\eta_x = 0.5$ and $\eta_\alpha = 0.2$.

The desired path for the vehicle was defined as $x_d = 300\sin(\theta)$ and $y_d = 300\cos(\theta)$, with $\theta(0) = 0$ and being updated by~\eqref{eq:th}. The desired surge speed was defined to obey the algorithm~\ref{alg1}. The functions defined in the algorithm are described in~\eqref{eq:sgn} and~\eqref{eq:interp}. The parameters for the algorithm are: $v_{xd,\text{min}} = 0.1$, $v_{xd,\text{max}} = 7$, $y_{e,\text{min}} = 1$, and $\tilde{\alpha}_{\text{min}} = 5^\circ$.

\begin{algorithm}[ht]
    Inputs: $v_{xd},~v_{xd,\text{min}},~v_{xd,\text{max}},~y_{e,\text{min}},~\tilde{\alpha}_{\text{min}}$\\
    \textbf{if} $\vert \tilde{\alpha} \vert < \tilde{\alpha}_{\text{min}}$ and $\vert y_e \vert < y_{e,\text{min}}$\\
    \hspace*{5mm} $\Delta v_{xd} \leftarrow 0$\\
    \textbf{else} \\
    \hspace*{5mm} \textbf{if} sgn($\tilde{\alpha}$) $=$ sgn($\dot{\tilde{\alpha}}$) \\
    \hspace*{10mm} $\Delta v_{xd} \leftarrow \text{interp}(\vert \tilde{\alpha} \vert, [\tilde{\alpha}_{\text{min}}, \pi/4], [-0.01,-0.1])$\\
    \hspace*{5mm} \textbf{else} \\
    \hspace*{10mm} $\Delta v_{xd} \leftarrow \text{interp}(\vert \tilde{\alpha} \vert, [\tilde{\alpha}_{\text{min}}, \pi/4], [0.01,0.1])$\\
    $v_{xd} \leftarrow v_{xd} + \Delta v_{xd}$ 
    \caption{Desired surge speed.}\label{alg1}
\end{algorithm}
\begin{figure*}[ht]
    \centering
    \includegraphics[width=0.7\textwidth]{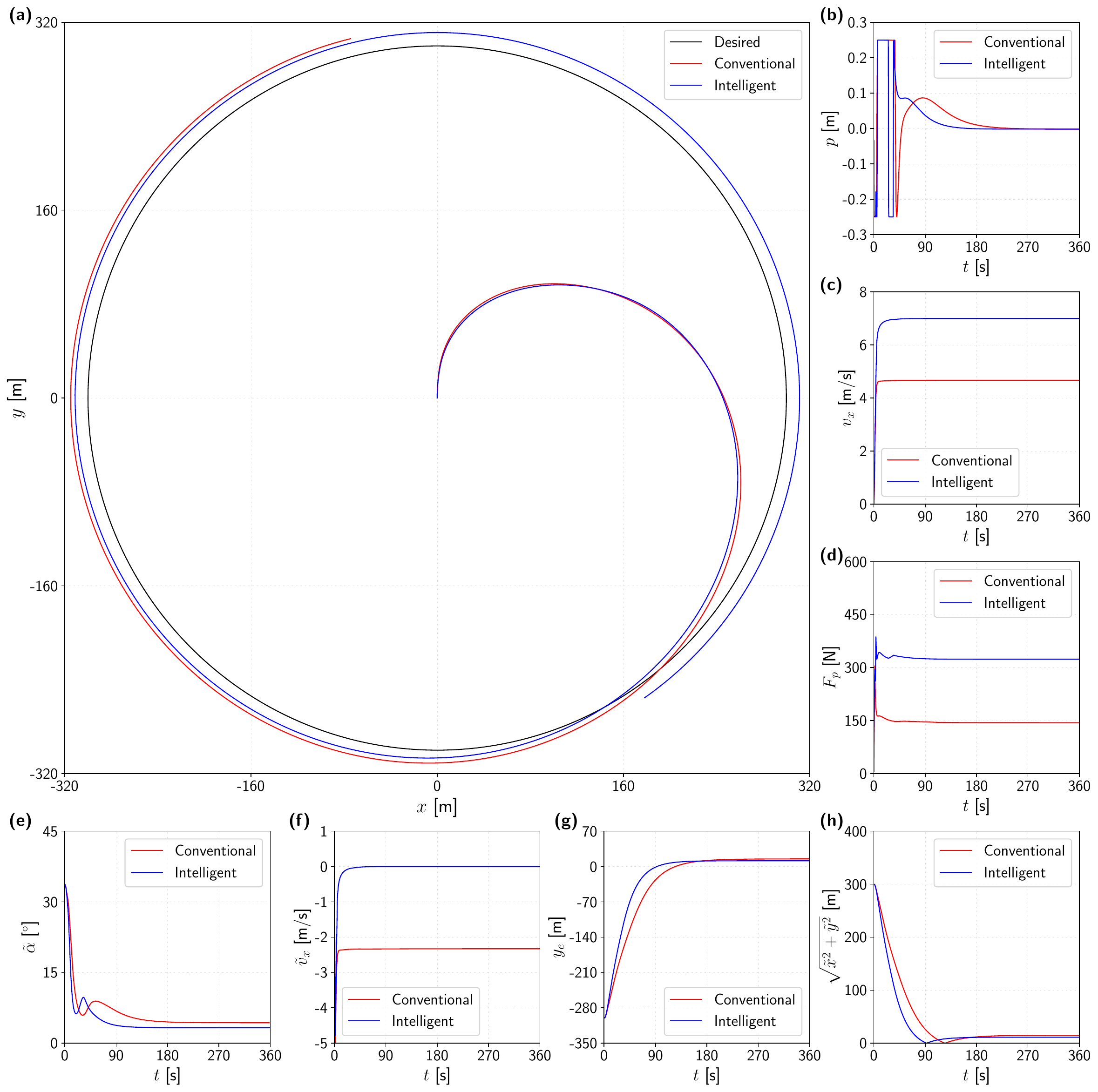}
    \caption{Simulation results considering $\dot{x}^\ast = 0$ and $\dot{y}^\ast = 0$: (a) Path-following performance; (b) sliding mass position; (c) surge speed; (d) thrust force; (e) heading angle error; (f) surge speed error; (g) cross-track error; and (h) absolute error.}\label{fig:r1}
\end{figure*}

The results shown in Fig. \ref{fig:r1} correspond to the case of calm water conditions, where $\dot{x}^\ast = 0$ and $\dot{y}^\ast = 0$. The intelligent approach is compared with the conventional controller (easily obtained by setting $\eta_x = \eta_\alpha = 0$). By comparison, we can see that the vehicle with the intelligent controller achieves a longer path than with the conventional approach. This is mainly due to the intelligent controller's ability to compensate for unknown dynamics and improve performance by reaching a higher speed as the angle error decreases. By computing the integral of the time-weighted absolute error (ITWAE), the conventional controller achieved 65.9\,m$\cdot$s, while the intelligent controller achieved 47.9\,m$\cdot$s, representing a 27\% decrease.

\begin{equation}\label{eq:sgn}
    \text{sgn}(x) = \left\lbrace \begin{array}{cl}
        \dfrac{x}{\vert x \vert} &\quad\text{if}~\vert x \vert > 0 \\
        0  &\quad\text{otherwise}
    \end{array} \right.
\end{equation}
\begin{multline}\label{eq:interp}
    \text{interp}(x, [x_{\text{min}}, x_{\text{max}}], [y_{\text{min}}, y_{\text{max}}]) = \\ \left\lbrace \begin{array}{cl}
        y_{\text{min}} & \quad\text{if}~x < x_{\text{min}} \\
        y_{\text{max}} & \quad\text{if}~x > x_{\text{max}} \\
        y_{\text{min}} + \tan\left( \dfrac{y_{\text{min}} - y_{\text{max}}}{x_{\text{min}} - x_{\text{max}}} \right) (x - x_{\text{min}})  &\quad\text{otherwise}
    \end{array} \right.
\end{multline}
\begin{figure*}[ht]
    \centering
    \includegraphics[width=0.7\textwidth]{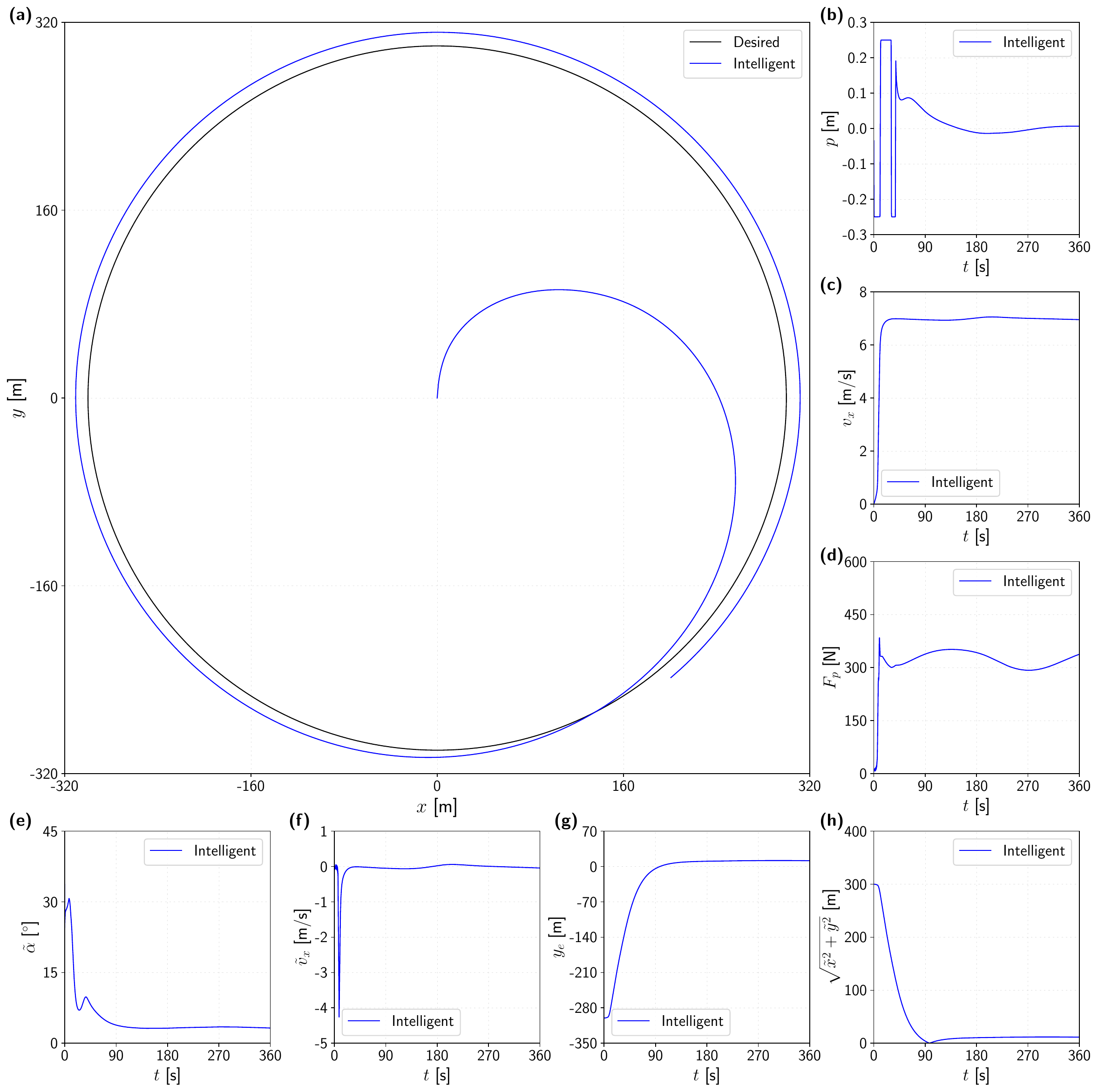}
    \caption{Simulation results considering $\dot{x}^\ast = 0.3$\,m/s and $\dot{y}^\ast = -0.2$\,m/s: (a) Path-following performance; (b) sliding mass position; (c) surge speed; (d) thrust force; (e) heading angle error; (f) surge speed error; (g) cross-track error; and (h) absolute error.}\label{fig:r2}
\end{figure*}

The results shown in the Fig. \ref{fig:r2} are for the case when the water stream has constant velocity of $\dot{x}^\ast = 0.3$\,m/s and $\dot{y}^\ast = -0.2$\,m/s. It is possible to see that the intelligent controller keeps the vehicle following the path with small error. As a matter of fact, the ITWAE for this case was approximately 50.9\,m$\cdot$s, a value very close with the previous case. Figure \ref{fig:r2}b also shows how the sliding mass position changes to compensate for the effects of the water conditions.

\begin{figure*}[ht]
    \centering
    \includegraphics[width=0.7\textwidth]{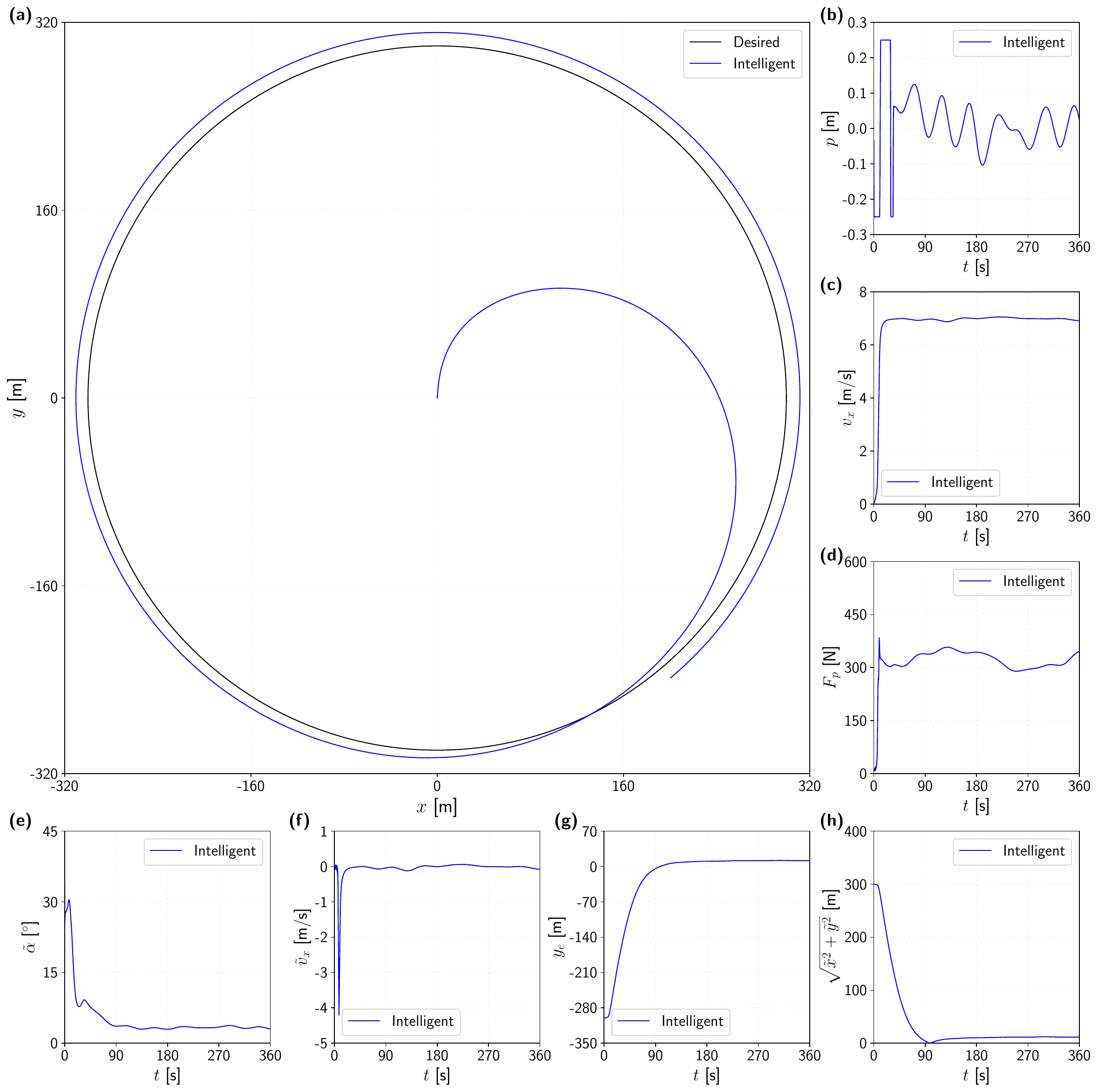}
    \caption{Simulation results considering $\dot{x}^\ast = 0.3 + 0.1\sin(0.11t)$~m/s and $\dot{y}^\ast = -0.2 + 0.1\sin(0.09t)$~m/s: (a) Path-following performance; (b) sliding mass position; (c) surge speed; (d) thrust force; (e) heading angle error; (f) surge speed error; (g) cross-track error; and (h) absolute error.}\label{fig:r3}
\end{figure*}

Finally, we have the results shown in the Fig. \ref{fig:r3} for the case when the water stream varies over the time: $\dot{x}^\ast = 0.3 + 0.1\sin(0.11t)$\,m/s and $\dot{y}^\ast = -0.2 + 0.1\sin(0.09t)$\,m/s. The intelligent controller effectively keeps the vehicle on the path with minimal error. In fact, the ITWAE for this scenario was approximately 50.8\,m$\cdot$s, a value very close to that of the first case. Additionally, Fig. \ref{fig:r3}b shows how the sliding mass position changes over time to compensate for the interference of water conditions.

\section{CONCLUSIONS}

We presented an intelligent controller for a surface vehicle maneuvered by a propeller and a dynamically adjustable sliding mass to change the vessel’s direction. We utilized a LOS algorithm to generate the desired states, while artificial neural networks approximate and compensate for unknown dynamics and external perturbations. Simulation results demonstrated the efficacy of the proposed controller in solving the path-following problem for surface vehicles subject to dynamic changes in mass distribution.

\addtolength{\textheight}{-0cm}   







\bibliographystyle{IEEEtran}
\bibliography{board}

\end{document}